\documentclass[a4paper,12pt]{article}
\pdfoutput=1

\usepackage[utf8]{inputenc}
\usepackage[english]{babel}
\usepackage{xspace} 
\usepackage{amsmath,amssymb,amsfonts,amsthm}

\usepackage{siunitx}
\DeclareSIUnit\particle{particle} % pour définir de nouveaux symboles
\DeclareSIUnit\stiffness{\pico\newton/\micro\meter}
\DeclareSIUnit\kbT{$k_{\text{B}} T$}

\usepackage{graphicx}
\usepackage{caption}
\usepackage{subcaption} % permet de faire des subfigures (remplace le package subfig)

\usepackage{booktabs}
\usepackage{array}
\usepackage{paralist} % si besoin regarder "enumitem" qui est un peu différent

\usepackage{textcomp} % rajoute des symboles (notes de musiques, etc.)
\usepackage{xcolor} % pour ajouter de la couleur (si besoin)
\usepackage{cite} % pour faire citations de façon pratique (notamment mettre [1-3] au lieu de [1,2,3]).

\usepackage{hyperref}
\usepackage{bookmark}

\definecolor{linkcolor}{rgb}{0,0,0.6} % couleur des liens (bleu foncé)
\hypersetup{
	colorlinks=true, % colore les liens au lieu de les encadrer
	pdfstartview=FitV, % ouvre le PDF de façon à ce qu'il prenne la taille verticale de l'écran
	urlcolor=linkcolor, % choix de la couleur des liens URL
	linkcolor= linkcolor, % choix de la couleur des liens internes (table des matières, etc.)
	citecolor=linkcolor % choix de la couleur des liens de citations
}

\begin{document}
\title{\bf {Information and thermodynamics: \\ Experimental verification  \\ of Landauer's erasure principle}}
\author{Antoine B\'{e}rut, Artyom Petrosyan and Sergio Ciliberto  \\   \\ Universit\'e de Lyon, 
Ecole Normale Sup\'erieure de Lyon,\\ Laboratoire de Physique ,
C.N.R.S. UMR5672,  \\ 46, All\'ee d'Italie, 69364 Lyon,  France\\}
\maketitle

\begin{abstract}
{ We present an experiment in which a one-bit  memory is constructed, using a system of a single colloidal particle trapped in a modulated 
double-well potential. We measure the amount of heat dissipated to erase a bit and we establish that in the limit of long erasure cycles the mean
dissipated heat saturates at the Landauer bound, i.e. the  minimal quantity of heat  necessarily produced to delete a  classical bit of 
information. This result demonstrates the intimate link between information theory and thermodynamics. To stress this connection we also show 
that a detailed Jarzynski equality is verified, retrieving the Landauer's bound  independently of the work done on the system. The experimental 
details  are presented and the experimental errors  carefully discussed}
\end{abstract}

\tableofcontents

\section{A link between information theory and $ \, $ thermodynamics}

The Landauer's principle was first introduced by Rolf Landauer in 1961~\cite{Landauer1961}. It states that any logically irreversible transformation 
of classical information is necessarily accompanied by the dissipation of at least $k_{\text{B}}T \ln 2$ of heat per lost bit, where $k_{\text{B}}$ is 
the Boltzmann constant and $T$ is the temperature. This quantity represents only $\sim \SI{3e-21} {\joule }$ at room temperature ($\SI{300} {\kelvin}$) 
but is a general lower bound, independent of the specific kind of memory system used.

An operation is said to be logically irreversible if its input cannot be uniquely determined from its output. Any Boolean function that maps several 
input states onto the same output state, such as AND, NAND, OR and XOR, is therefore logically irreversible. In particular, the erasure of 
information, the RESET TO ZERO operation, is logically irreversible and leads to an entropy increase of at least $k_{\text{B}} \ln 2$ per erased 
bit.

A simple example can be done with a 1-bit memory system (\textit{i.e.} a systems with two states, called 0 and 1) modelled by a physical double 
well potential in contact with a single heat bath. In the initial state, the bistable potential is considered to be at equilibrium with the heat bath, and 
each state (0 and 1) have same probability to occur. Thus, the entropy of the system is $S = k_{\text{B}} \ln 2$, because there are two states with 
probability $1/2$. If a RESET TO ZERO operation is applied, the system is forced into state 0. Hence, there is only one accessible state with 
probability 1, and the entropy vanishes $S = 0$. Since the Second Law of Thermodynamics states that the entropy of a closed system cannot 
decrease on average, the entropy of the heat bath must increase of at least $k_{\text{B}} \ln 2$ to compensate the memory system's loss of 
entropy. This increase of entropy can only be done by an heating effect: the system must release in the heat bath at least $k_{\text{B}}T \ln 2$ of 
heat per bit erased\footnote{It it sometimes stated that the cost is $k_{\text{B}}T \ln 2$ per bit written. It is actually the same operation as the 
RESET TO ZERO can also be seen to store one given state (here state 0), starting with an unknown state.}.

For a reset operation with efficiency smaller than 1 (\textit{i.e.} if the operation only erase the information with a probability $p<1$), the Landauer's 
bound is generalised:
\begin{equation}
\langle Q \rangle \geq k_{\text{B}}T \left[ \ln 2 + p \ln (p) + (1-p) \ln (1-p) \right]
\label{landauer:eq:landauer_tot}
\end{equation}

The Landauer's principle was widely discussed as it could solve the paradox of Maxwell's ``demon''~\cite
{Brillouin1956,Penrose1970,Bennett1982}. The demon is an intelligent creature able to monitor individual molecules of a gas contained in two 
neighbouring chambers initially at the same temperature. Some of the molecules will be going faster than average and some will be going slower. 
By opening and closing a molecular-sized trap door in the partitioning wall, the demon collects the faster (hot) molecules in one of the chambers 
and the slower (cold) ones in the other. The temperature difference thus created can be used to run a heat engine, and produce useful work. By 
converting information (about the position and velocity of each particle) into energy, the demon is therefore able to decrease the entropy of the 
system without performing any work himself, in apparent violation of the Second Law of Thermodynamics. A simpler version with a single particle, 
called Szilard Engine~\cite{Szilard1929} has recently been realised experimentally~\cite{Toyabe2010}, showing that information can indeed be 
used to extract work from a single heat bath. The paradox can be resolved by noting that during a full thermodynamic cycle, the memory of the 
demon, which is used to record the coordinates of each molecule, has to be reset to its initial state. Thus, the energy cost to manipulate the 
demon's memory compensate the energy gain done by sorting the gas molecules, and the Second Law of Thermodynamics is not violated any 
more.

More information can be found in the two books~\cite{Leff1990,Leff2002}, and in the very recent review~\cite{Parrondo2015} about 
thermodynamics of information based on stochastic thermodynamics and fluctuation theorems.

In this article, we describe an experimental realization of the Landauer's information erasure procedure, using a Brownian particle trapped with 
optical tweezers in a time-dependent double well potential. This kind of system was theoretically~\cite{Shizume1995} and numerically~\cite
{Lutz2009} proved to show the Landauer's bound $k_{\text{B}}T \ln 2$ for the mean dissipated heat when an information erasure operation is 
applied. The results described in this article were partially presented in two previous  articles~\cite{Berut2012,Berut2013}, and were later 
confirmed by two independent experimental works~\cite{Roldan2014,Bechhoefer2014}. 

{{The article is
organized as follows. In section 2 we describe  the experimental set-up and the one bit memory. 
In section 3 the experimental results on the Landauer's bound and the dissipated heat are  presented. 
In section 4 we analyze the experimental results within the context of the  Jarzinsky equality.
 Finally we conclude in section 5.}}

\section{Experimental set-up}

\subsection{The one-bit memory system}

\begin{figure}%[ht!]
\begin{center}
\includegraphics[width=6cm]{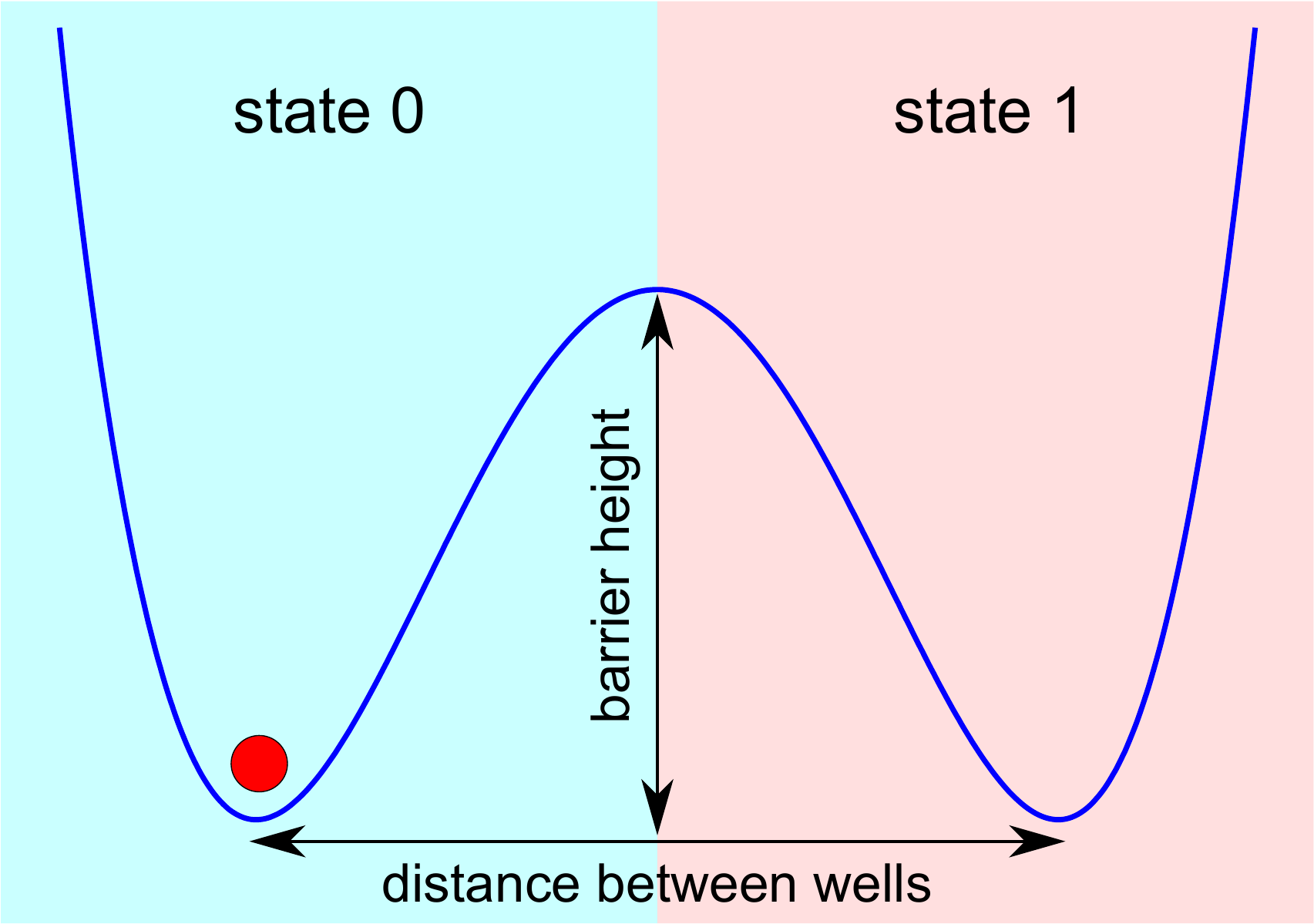}
\caption{Schematic representation of the one-bit memory system, made of one particle trapped in a double well potential.}
 \label{landauer:fig:1bit}
\end{center}
\end{figure}

The one-bit memory system, is made of a double well potential where one particle is trapped by optical tweezers. If the particle is in the left-well 
the system is in the state ``0'', if the particle is in the right-well the system in the state ``1'' ( see fig.~\ref{landauer:fig:1bit}).

\begin{figure}[ht!]
\begin{center}
\includegraphics[width=10cm]{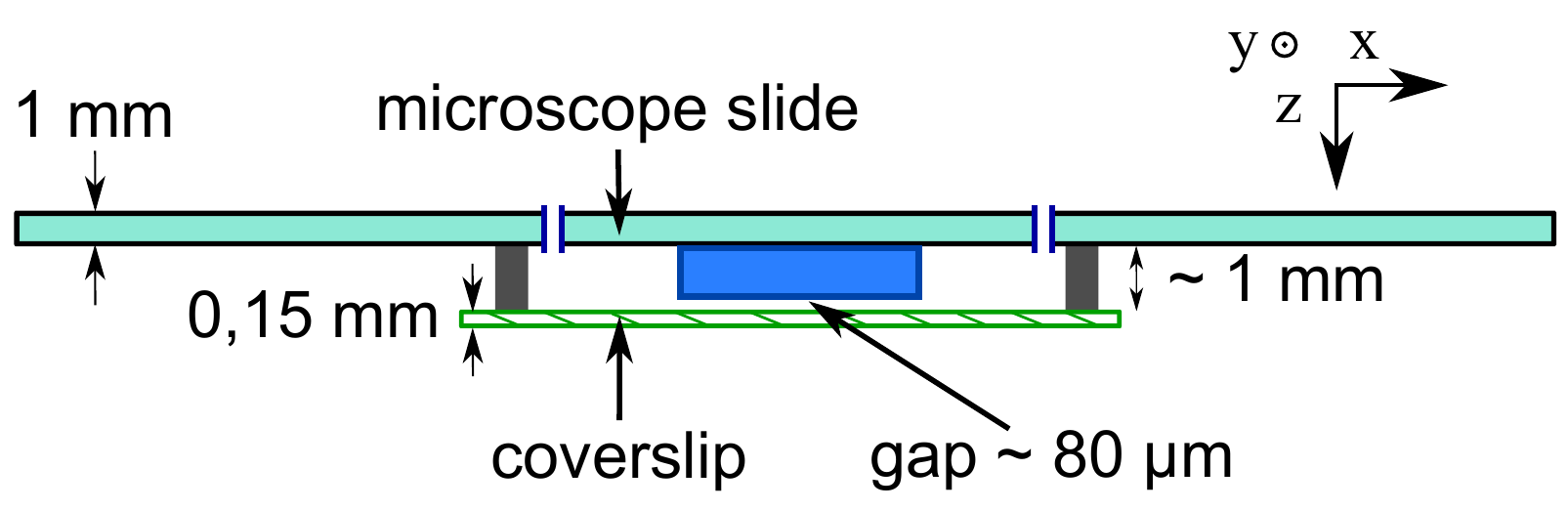}
\caption{Schematic representation of the cell used to trap particles dispersed in water (view from the side). The central part has a smaller gap 
than the rest of the cell.} 
\label{landauer:fig:cell}
\end{center}
\end{figure}

The particles are silica beads (radius $R = \SI{1.00(5)} {\micro\meter}$), diluted a low concentration in bidistilled water.
 The solution is contained in 
a disk-shape cell. The center of the cell has a smaller depth ($\sim \SI{80} {\micro\meter}$) compared to the rest of the cell 
($\sim \SI{1}  {\milli \meter}$), see figure~\ref{landauer:fig:cell}. 
This central area contains less particles than the rest of the cell and provides us a clean region where 
one particle can be trapped for a long time without interacting with other particles.

\begin{figure}[ht!]
\begin{center}
\includegraphics[width=13cm]{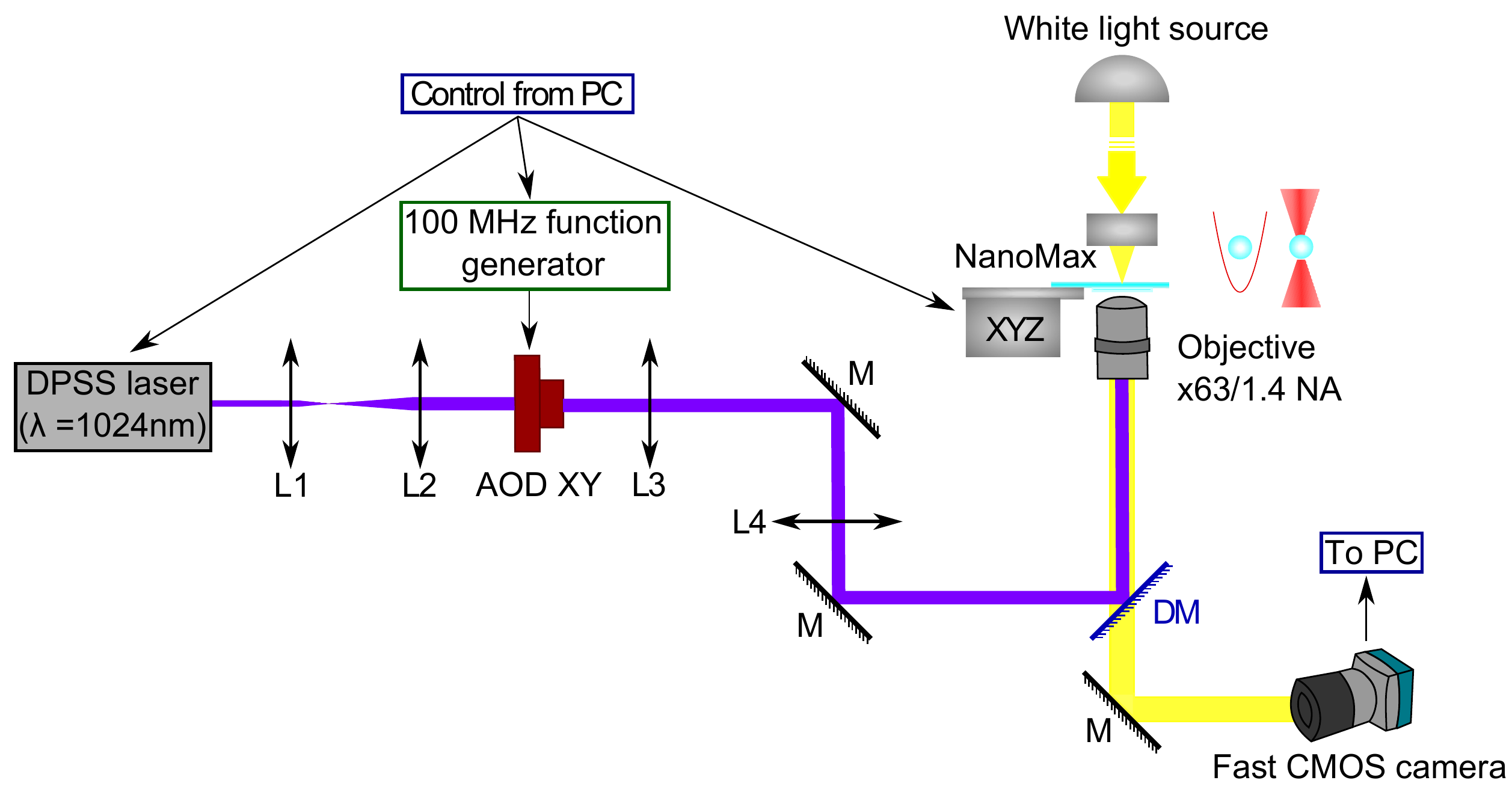}
\caption{Schematic representation of optical tweezers set-up used to trap one particle in a double well potential. 
The Acousto-Optic Deflector 
(AOD) is used to switch rapidly the trap between two positions. The NanoMax piezo stage can move the cell with regard to the laser, which 
creates a flow around the trapped particle. ``M'' are mirrors and ``DM'' is a dichroic mirror.} 
\label{landauer:fig:setup_laser}
\end{center}
\end{figure}

The double well potential is created using an Acousto-Optic Deflector which allows us to switch very rapidly (at a rate of $\SI{10} {\kilo\hertz}$) 
a 
laser beam (wavelength $\lambda = \SI{1024} {\nano\meter}$) between two positions (separated by a fixed distance 
$d \sim \SI{1}{\micro\meter}$). These two positions become for the particle the two wells of the double well potential. The intensity of the laser
 $I$ can be controlled from 
 $\SI {10} {\milli\watt}$ to more than $\SI{100} {\milli\watt}$, 
which enables us to change the height of the double well potential's central barrier \footnote{The 
values are the power measured on the beam before the microscope objective, so the ``real'' power at the focal point should be smaller, due to the 
loss in the objective.}. A NanoMax closed-loop piezoelectric stage from Thorlabs$^{\circledR}$ with high resolution ($\SI{5} {\nano\meter}$) 
can 
move the cell with regard to the position of the laser. Thus it allows us to create a fluid flow around the trapped particle. The position of the bead is 
tracked using a fast
camera with a resolution of $\SI{108} {\nano\meter}$ per pixel, which after treatment gives the position with a precision greater than
$\SI{5} {\nano\meter}$. The trajectories of the bead are sampled at 502 Hz. See figure~\ref{landauer:fig:setup_laser}.

\begin{figure}[ht!]
\begin{center}
\includegraphics[width=9cm]{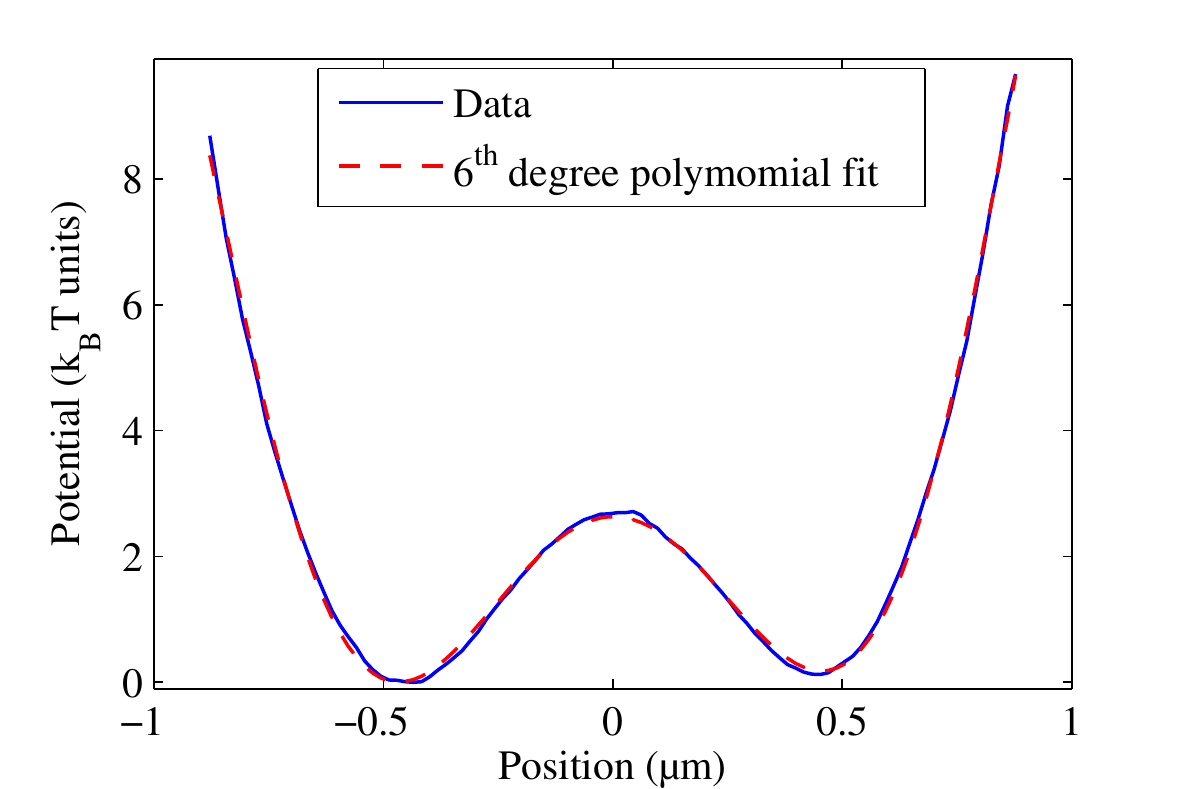}
\caption{Double well potential measured by computing the equilibrium distribution of one particle's positions, with 
$I_{\text{laser}} = \SI{15}{\milli\watt}$. Here the distribution is computed on ${1.5e6}$ points sampled at 
$\SI{502}{\hertz} (\textit{i.e.} a \SI{50} {\minute} $long measurement). 
The double well potential is well fitted by a $6^{th}$ degree polynomium.} 
\label{landauer:fig:potentiel_repos}
\end{center}
\end{figure}

The beads are trapped at a distance $h = \SI{25} {\micro\meter}$ from the bottom of the cell. The double well potential must be tuned for each 
particle, in order to be as symmetrical as possible and to have the desired central barrier. 
The tuning is done by adjusting the distance between the two traps and the time that the laser spend on each trap. 
The asymmetry can be reduced to $ \sim \SI{0.1}   {\kbT} $. 
The double well potential $U_0(x,I)$ (with $x$ the position and $I$ the intensity of the laser) can simply be measured by computing the 
equilibrium distribution of the position for one particle in the potential:
\begin{equation}
P(x,I) \propto \exp(-\frac{U(x,I)}{k_{\text{B}}T}).
\end{equation}
One typical double well potential is shown in figure~\ref{landauer:fig:potentiel_repos}.

\subsection{The information erasure procedure}
\label{landaeur:sec:description_erasure}

\begin{figure}[ht!]
        \centering
        \begin{subfigure}[c]{0.5\textwidth}
                \includegraphics[width=\textwidth]{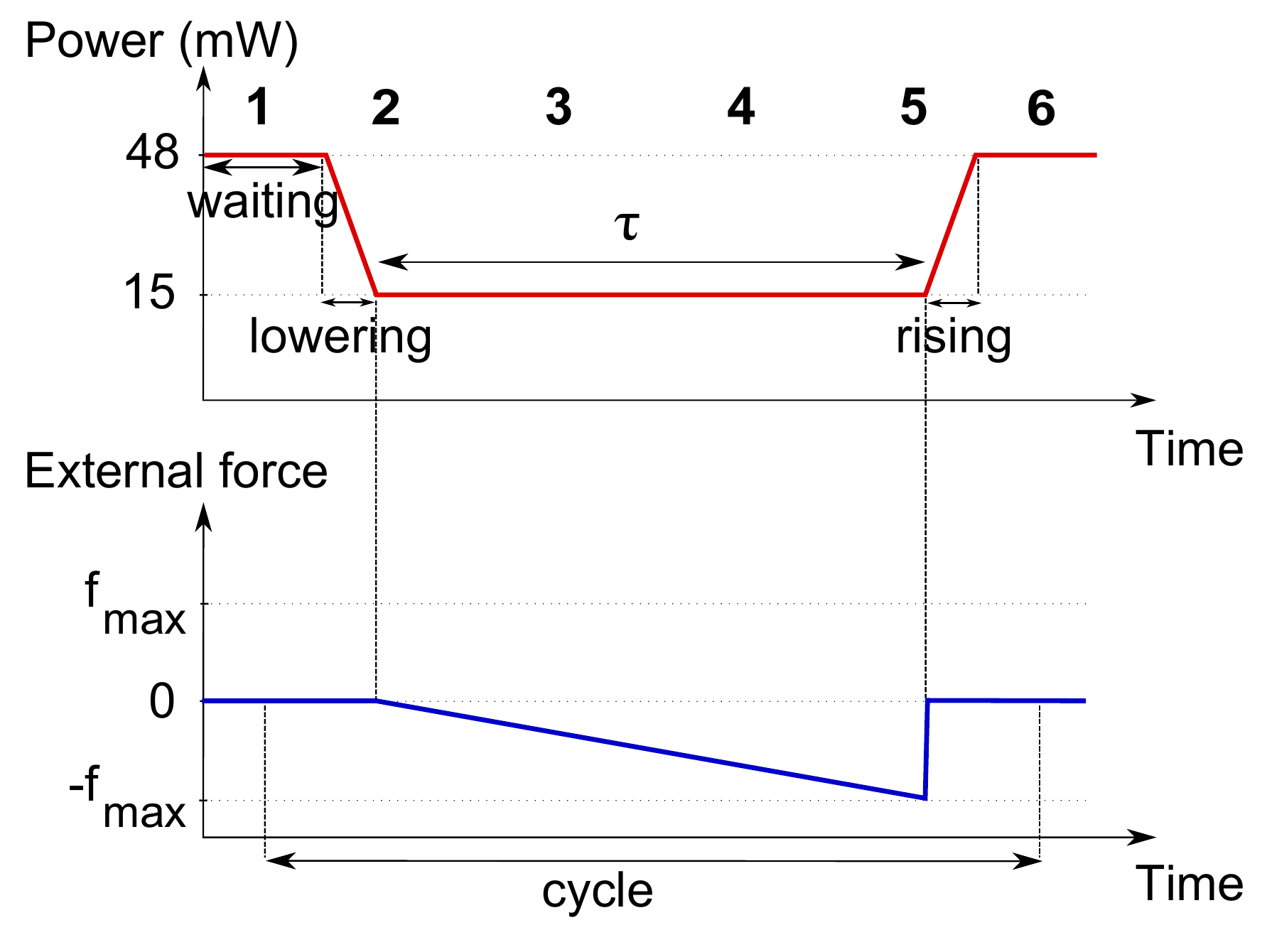}
                \caption{Laser intensity and external drag force.}
                \label{landauer:fig:procedure}
        \end{subfigure}%
        \begin{subfigure}[c]{0.5\textwidth}
                \includegraphics[width=\textwidth]{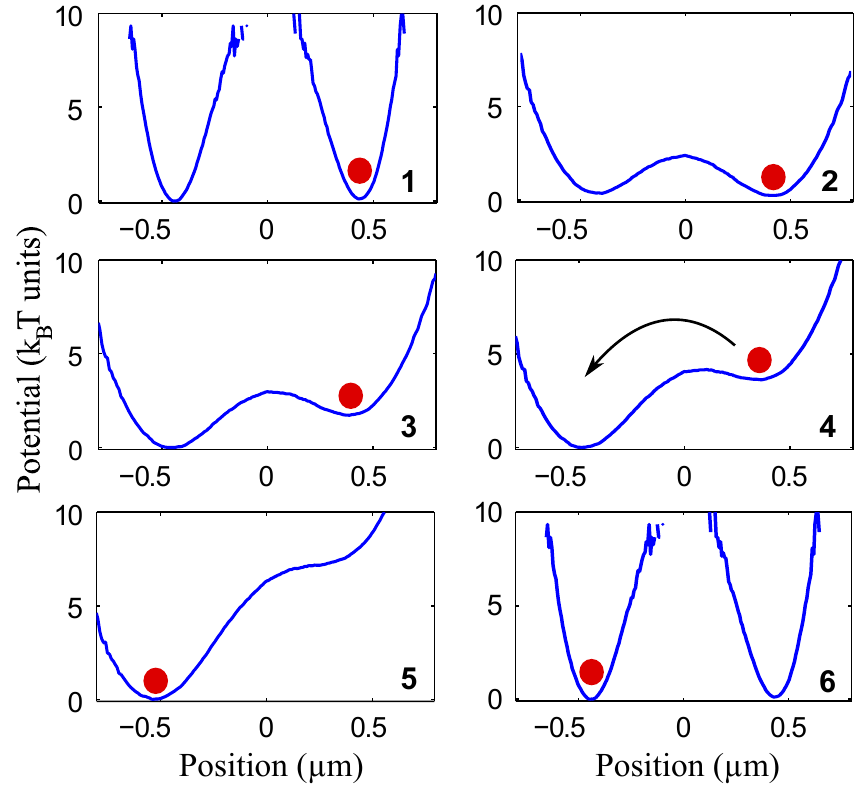}
                \caption{Potential felt by the particle.}
                \label{landauer:fig:potential}
        \end{subfigure}
        \caption{Schematical representation of the erasure procedure. The potential felt by the trapped particle is represented at different stages of 
the procedure (\textbf{1} to \textbf{6}). For \textbf{1} and \textbf{2} the potential $U_0(x,I)$ is measured. For \textbf{3} to \textbf{5} the potential is 
constructed from $U_0(x,I_{\text{low}})$ knowing the value of the applied drag force.}
        \label{landauer:fig:erasure}
\end{figure}

We perform the erasure procedure as a logically irreversible operation. This procedure brings the system initially in one unknown state (0 or 1 
with same probability) in one chosen state (we choose 0 here). It is done experimentally in the following way (and summarised in figure~\ref
{landauer:fig:procedure}):

\begin{itemize}
\item At the beginning the bead must be trapped in one well-defined state (0 or 1). For this reason, we start with a high laser intensity 
($I_{\text{high}}= \SI{48}{\milli\watt}$) so that the central barrier is more than $\SI{8}  {\kbT}$. In this situation, the characteristic jumping time (Kramers Time) 
is about \SI{3000}  {\second}, which is long compared to the time of the experiment, and the equivalent stiffness of each well is about \SI{1.5}
{\stiffness}. The system is left \SI{4}  {\second} with high laser intensity so that the bead is at equilibrium in the well where it is trapped 
\footnote{The characteristic time for the particle trapped in one well when the barrier is high is $\SI{0.08}  {\second}$}. The potential $U_0(x,I_{\text{high}})$ 
is represented in figure~\ref{landauer:fig:potential} \textbf{1}.
\item The laser intensity is first lowered (in a time $T_{\text{low}} = \SI{1}  {\second}$) to a low value ($I_{\text{low}} =\SI{15}{\milli\watt}$) so that 
the barrier is about $\SI{2.2}  {\kbT}$. In this situation the jumping time falls to
 $\sim \SI{10}  {\second}$, and the equivalent stiffness of each well is 
about $\SI{0.3}{\stiffness}$. The potential $U_0(x,I_{\text{low}})$ is represented figure~\ref{landauer:fig:potential} \textbf{2}.
\item A viscous drag force linear in time is induced by displacing the cell with respect to the laser using the piezoelectric stage. The force is given 
by $f= \gamma v$ where $\gamma=6 \pi R \eta$ ($\eta$ is the viscosity of water) and $v$ the speed of displacement. It tilts the double well 
potential so that the bead ends always in the same well (state 0 here) independently of where it started. See figures~\ref{landauer:fig:potential} 
\textbf{3} to \textbf{5}.
\item At the end, the force is stopped and the central barrier is raised again to its maximal value (in  a time $T_{\text{high}} = \SI{1}  {\second}$). 
See figure~\ref{landauer:fig:potential} \textbf{6}.
\end{itemize}

The total duration of the erasure procedure is $T_{\text{low}}+\tau+T_{\text{high}}$. Since we kept 
$T_{\text{low}} = T_{\text{high}} = \SI{1}  {\second}$, a procedure is fully characterised by the duration $\tau$ and the maximum value of the force applied $f_{\text{max}}$. Its efficiency is 
characterized by the ``proportion of success'' $P_S$, which is the proportion of trajectories where the bead ends in the chosen well (state 0), 
independently of where it started.

\begin{figure}[ht!]
        \centering
        \begin{subfigure}[c]{0.5\textwidth}
                \includegraphics[width=\textwidth]{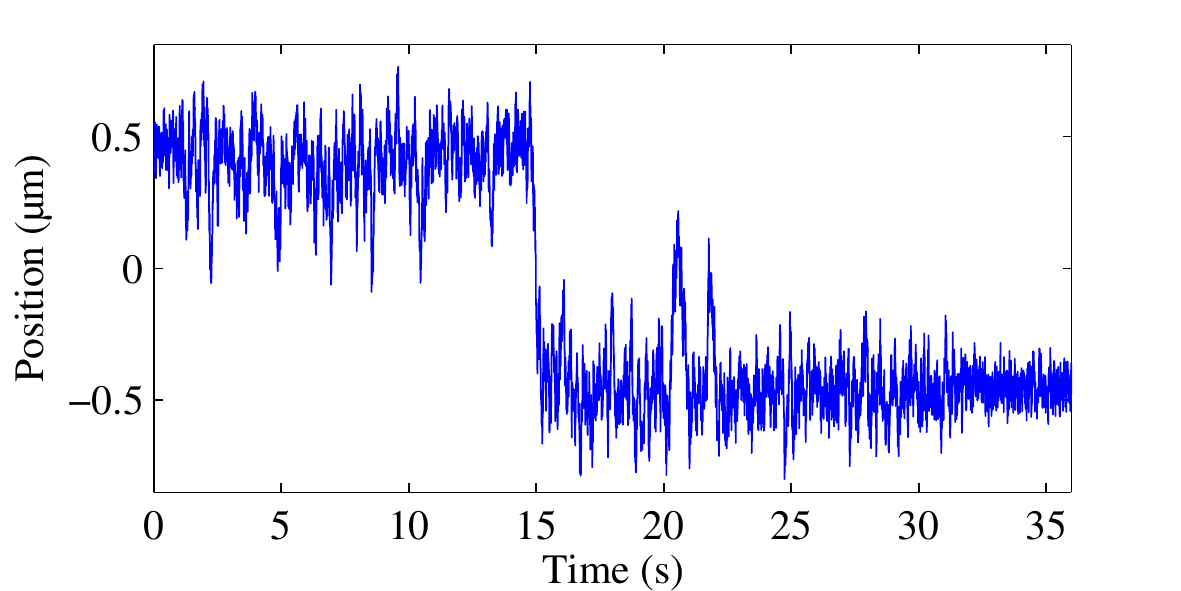}
                \caption{$1 \rightarrow 0$}
                \label{landauer:fig:1vers0}
        \end{subfigure}%
        \begin{subfigure}[c]{0.5\textwidth}
                \includegraphics[width=\textwidth]{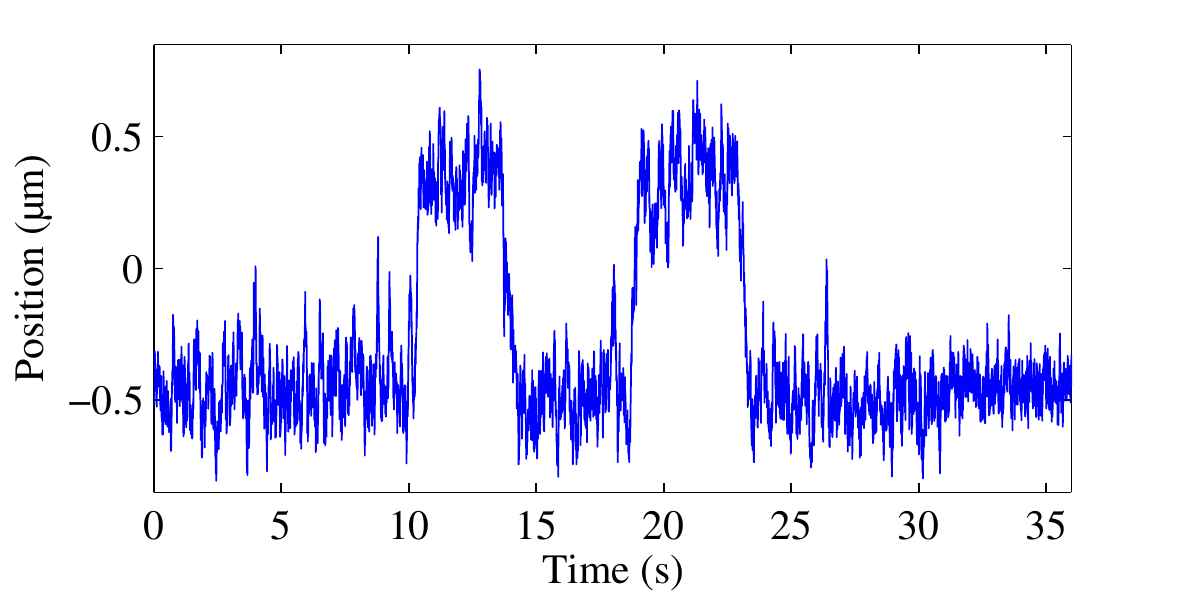}
                \caption{$0 \rightarrow 0$}
                \label{landauer:fig:0vers0}
        \end{subfigure}
        \caption{Examples of trajectories for the erasure procedure. $t=0$ corresponds to the time where the barrier starts to be lowered. The two 
possibilities of initial state are shown.}
        \label{landauer:fig:exemple_trajectoires}
\end{figure}

Note that for the theoretical procedure, the system must be prepared in an equilibrium state with same probability to be in state 1 than in state 0. 
However, it is more convenient experimentally to have a procedure always starting in the same position. Therefore we separate the procedure in 
two sub-procedures: one where the bead starts in state 1 and is erased in state 0, and one where the bead starts in state 0 and is erased in state 
0. The fact that the position of the bead at the beginning of each procedure is actually known is not a problem because this knowledge is not used 
by the erasure procedure. The important points are that there is as many procedures starting in state 0 than in state 1, and that the procedure is 
always the same regardless of the initial position of the bead. Examples of trajectories for the two sub-procedures $1 \rightarrow 0$ and $0 
\rightarrow 0$ are shown in figure~\ref{landauer:fig:exemple_trajectoires}.

\newpage

\section{Landauer's bound for dissipated heat}

\subsection{Computing the dissipated heat}

The system can be described by an over-damped Langevin equation:
\begin{equation}
\gamma \dot{x} = - \frac{\partial U_0}{\partial x}(x,I) + f(t) + \xi(t)
\label{landauer:eq:langevin}
\end{equation}
with $x$ the position of the particle\footnote{We take the reference $x=0$ as the middle position between the two traps.}, $\dot{x} = \frac{\mathrm
{d}x}{\mathrm{d}t}$ its velocity, $\gamma = 6 \pi R \eta$ the friction coefficient ($\eta$ is the viscosity of water), $U_0(x,I)$ the double well 
potential created by the optical tweezers, $f(t)$ the external drag force exerted by displacing the cell, and $\xi(t)$ the thermal noise which verifies 
$\left\langle \xi(t) \right\rangle = 0$ and $\left\langle \xi(t)\xi(t') \right\rangle =2 \gamma k_{\text{B}} T \delta (t-t')$, where $\left\langle . \right\rangle$ 
stands for the ensemble average.

Following the formalism of the stochastic energetics~\cite{Sekimoto1998}, the heat dissipated by the system into the heat bath along the 
trajectory $x(t)$ between time $t=0$ and $t$ is:
\begin{equation}
Q_{0,t} = \int_{0}^{t} - \left( \xi(t^{\prime}) - \gamma \dot{x}(t^{\prime})  \right) \dot{x}(t^{\prime}) \, \mathrm{d} t^{\prime}.
\end{equation}
Using equation~\ref{landauer:eq:langevin}, we get:
\begin{equation}
Q_{0,t} = \int_{0}^{t} \left( - \frac{\partial U_0}{\partial x}(x,I) + f(t^{\prime}) \right) \dot{x} (t^{\prime}) \, \mathrm{d} t^{\prime}.
\end{equation}
For the erasure procedure described in~\ref{landaeur:sec:description_erasure} the dissipated heat can be decomposed in three terms:
\begin{equation}
Q_{\text{erasure}} = Q_{\text{barrier}} + Q_{\text{potential}} + Q_{\text{drag}}
\end{equation}
Where:
\begin{itemize}
\item $Q_{\text{barrier}}$ is the heat dissipated when the central barrier is lowered and risen ($f = 0$ during these stages of the procedure):
\begin{equation}
Q_{\text{barrier}} = \int_{0}^{T_{\text{low}}} \left( - \frac{\partial U_0}{\partial x}(x,I) \right) \dot{x} \, \mathrm{d} t^{\prime} + \int_{T_{\text{low}}+\tau}^
{T_{\text{low}}+\tau+T_{\text{high}}} \left( - \frac{\partial U_0}{\partial x}(x,I) \right) \dot{x} \, \mathrm{d} t^{\prime}
\end{equation}
\item $Q_{\text{potential}}$ is the heat dissipated due to the force of the potential, during the time $\tau$ where the external drag force is applied 
(the laser intensity is constant during this stage of the procedure):
\begin{equation}
Q_{\text{potential}} = \int_{T_{\text{low}}}^{T_{\text{low}}+\tau} \left( - \frac{\partial U_0}{\partial x}(x,I) \right) \dot{x} \, \mathrm{d} t^{\prime}
\end{equation}
\item $Q_{\text{drag}}$ is the heat dissipated due the external drag force applied during the time $\tau$ (the laser intensity is constant during this 
stage of the procedure):
\begin{equation}
Q_{\text{drag}} = \int_{T_{\text{low}}}^{T_{\text{low}}+\tau} f \dot{x} \, \mathrm{d} t^{\prime}
\end{equation}
\end{itemize}

The lowering and rising of the barrier are done in a time much longer than the relaxation time of the particle in the trap
 ($\sim \SI{0.1}  {\second}$), so they can be considered as a quasi-static cyclic process, and do not contribute to the dissipated heat in average. 
The complete calculation can also be done if we assume that the particle do not jump out of the well where it is during the change of barrier 
height. In this case, we can do a quadratic approximation: $U_0(x,I) = - k(I) x$ where $k$ is the stiffness of the trap, which evolves in time 
because it depends linearly on the intensity of the laser. Then:
\begin{equation}
\langle Q_{\text{lowering}} \rangle = \left\langle \int_{0}^{T_{\text{low}}} - k x \dot{x} \, \mathrm{d} t^{\prime} \right\rangle = \int_{0}^{T_{\text{low}}} - 
\frac{k}{2} \, \mathrm{d} \left( \langle x^2 \rangle \right).
\end{equation}
Using the equipartition theorem (which is possible because the change of stiffness is assumed to be quasi-static), we get:
\begin{equation}
\langle Q_{\text{lowering}} \rangle = \int_{0}^{T_{\text{low}}} - k \frac{k_{\text{B}}T}{2} \, \mathrm{d} \left( \frac{1}{k}  \right) = \frac{k_{B}T}{2} \ln \left
( \frac{k_{\text{low}}}{k_{\text{high}}} \right).
\end{equation}
The same calculation gives $\langle Q_{\text{rising}} \rangle = \frac{k_{B}T}{2} \ln \left( \frac{k_{\text{high}}}{k_{\text{low}}} \right)$, and it follows 
directly that $\langle Q_{\text{barrier}} \rangle = \langle Q_{\text{lowering}} \rangle  + \langle Q_{\text{rising}} \rangle =0 $.

The heat dissipated due to the potential when the force is applied is also zero in average. Indeed, the intensity is constant and $U_0$ becomes a 
function depending only on $x$. It follows that:
\begin{equation}
\langle Q_{\text{potential}} \rangle = \int_{T_{\text{low}}}^{T_{\text{low}}+\tau} - \frac{\mathrm{d} U_0}{\mathrm{d} x} \, \mathrm{d} x =  \Big[ U_0(x) 
\Big]_{x(T_{\text{low}}+\tau)}^{x(T_{\text{low}})}.
\end{equation}
Since the potential is symmetrical, there is no change in $U_0$ when the bead goes from one state to another, and $\left\langle U_0(x(T_{\text
{low}})) - U_0(x(T_{\text{low}}+\tau)) \right\rangle = 0$.

Finally, since we are interested in the mean dissipated heat, the only relevant term to calculate is the heat dissipated by the external drag force:
\begin{equation}
Q_{\text{drag}} = \int_{T_{\text{low}}}^{T_{\text{low}}+\tau} \gamma v(t^{\prime}) \dot{x} \, \mathrm{d} t^{\prime}.
\end{equation}
Where $\gamma$ is the known friction coefficient, $v(t)$ is the imposed displacement of the cell (which is not a fluctuating quantity) and $\dot{x}$ 
can be estimated simply:
\begin{equation}
\dot{x}(t+\delta t/2) = \frac{x(t+\delta t) - x(t)}{\delta t}.
\end{equation}
We measured $Q_{\text{drag}}$ for several erasure procedures with different parameters $\tau$ and $f_{\text{max}}$. For each set of 
parameters, we repeated the procedure a few hundred times in order to compute the average dissipated heat.

We didn't measure $Q_{\text{lowering}}$ and $Q_{\text{rising}}$ because it requires to know 
the exact shape of the potential at any time during lowering and rising of the central barrier. 
The potential could have been measured by computing the equilibrium distribution of one
 particle's positions for different values of $I$. But these measurements would have been very 
long since they require to be done on times much longer than the Kramers time to give 
a good estimation of the double well potential.  Nevertheless, 
we estimated on numerical simulations with parameters close to our experimental ones that 
$\langle Q_{\text{lowering}} + Q_{\text{potential}} + Q_{\text{rising}} \rangle \approx \SI{0.7}  {\kbT}$ 
which is only $\SI{10}{\percent}$ of the Landauer's bound.

\subsection{Results}

\begin{figure}[ht!]
\begin{center}
\includegraphics[width=9cm]{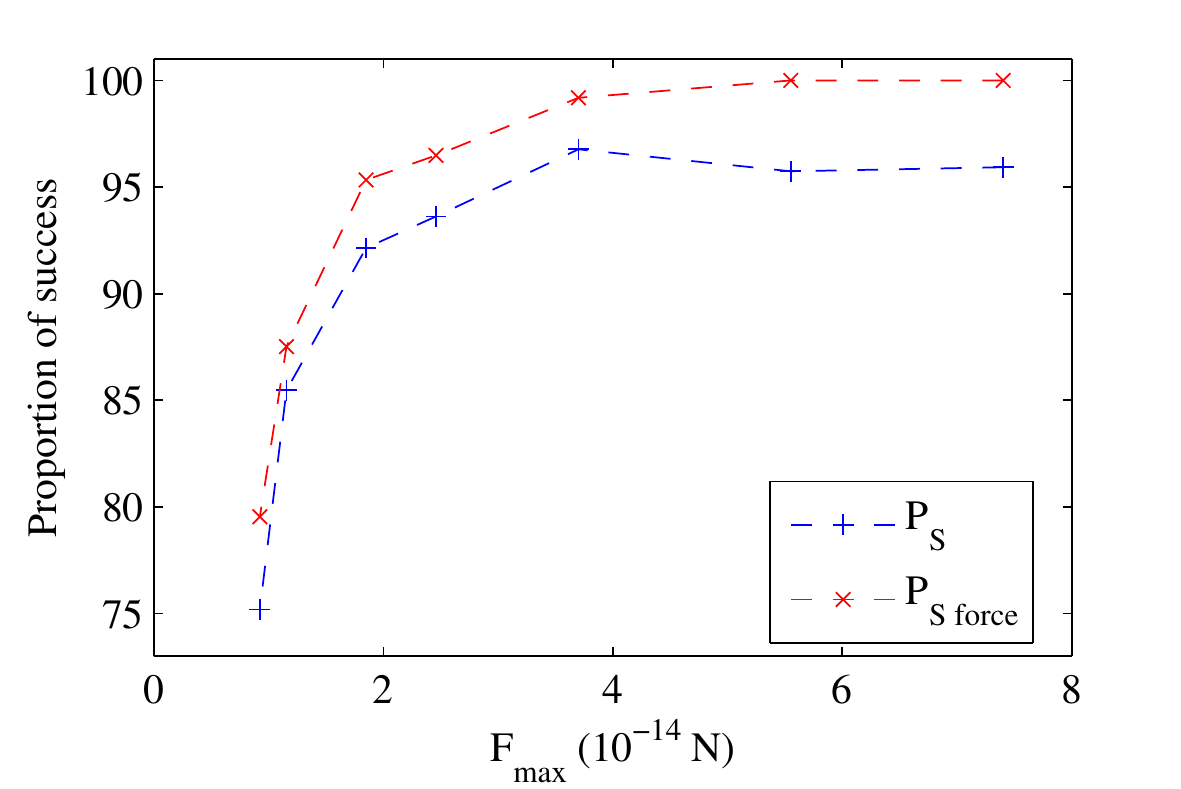}
\caption{Proportion of success for different values of $\tau$ and $f_{\text{max}}$, keeping constant the product 
$\tau \times  f_{\text{max}} \approx \SI{0.4}{\pico\newton\second}$ 
(which corresponds to $\tau \times v_{max} = \SI{20}{\micro\meter}$). $P_S$ quantifies the ratio of procedures 
which ends in state 0 after the barrier is risen. $P_{S \, \text{force}}$ quantifies the ratio of procedures which ends in state 0 before the barrier is 
risen.} 
\label{landauer:fig:probability_success}
\end{center}
\end{figure}

We first measured $P_S$ the proportion of success for different set of $\tau$ and $f_{\text{max}}$. 
`Qualitatively, the bead is more likely to jump from one state to another thanks to thermal fluctuations 
if the waiting time is longer. Of course it also has fewer chances to escape from state 0 if the force pushing
 it toward this state is stronger. We did some measurements keeping the product $\tau \times f_{\text{max}}$ constant. 
 The results are shown in figure~\ref{landauer:fig:probability_success} (blue points).

The proportion of success is clearly not constant when the product $\tau \times f_{\text{max}}$ is kept constant, but, as expected, the higher the 
force, the higher $P_S$. One must also note that the experimental procedure never reaches a $P_S$ higher than $\sim \SI{95}{\percent}$. This 
effect is due to the last part of the procedure: since the force is stopped when the barrier is low, the bead can always escape from state 0 during 
the time needed to rise the barrier. This problem can be overcome with a higher barrier or a faster rising time $T_{\text{high}}$. It was tested 
numerically by Raoul Dillenscheider and \'{E}ric Lutz, using a protocol adapted from~\cite{Lutz2009} to be close to our experimental procedure. 
They showed that for a high barrier of $\SI{8}  {\kbT}$ the proportion of success approaches only $\sim \SI{94}{\percent}$, whereas for a barrier of 
$\SI{15}  {\kbT}$ it reaches $\sim \SI{99}{\percent}$. Experimentally we define a proportion of success $P_{S \, \text{force}}$ by counting the number 
of procedures where the particle ends in state 0 when the force is stopped (\emph{before} the rising of the barrier). It quantifies the efficiency of 
the pushing force, which is the relevant one since we have shown that the pushing force is the only contribution to the mean dissipated heat. 
Measured $P_{S \, \text{force}}$ are shown in figure~\ref{landauer:fig:probability_success} (red points). $P_{S \, \text{force}}$ is roughly always 
$\SI{5}{\percent}$ bigger than $P_S$ and it reaches $\SI{100} {\percent}$ of success for high forces.

To reach the Landauer's bound, the force necessary to erase information must be as low as possible, because it is clear that a higher force will 
always produce more heat for the same proportion of success. Moreover, the bound is only reachable for a quasi-static (\textit{i.e.} $\tau 
\rightarrow \infty$) erasure procedure, and the irreversible heat dissipation associated with a finite time procedure should decrease as $1/\tau$~
\cite{SekimotoSasa1997}. Thus we decided to work with a chosen $\tau$ and to manually\footnote{The term ``manually'' refers to the fact that the 
optimisation was only empirical and that we did not computed the theoretical best $f_{\text{max}}$ for a given value of $\tau$.} optimise the 
applied force. The idea was to choose the lowest value of $f_{\text{max}}$ which gives a $P_{S \, \text{force}} \geq \SI{95}{\percent}$. 

\begin{figure}[ht!]
\begin{center}
\includegraphics[width=9cm]{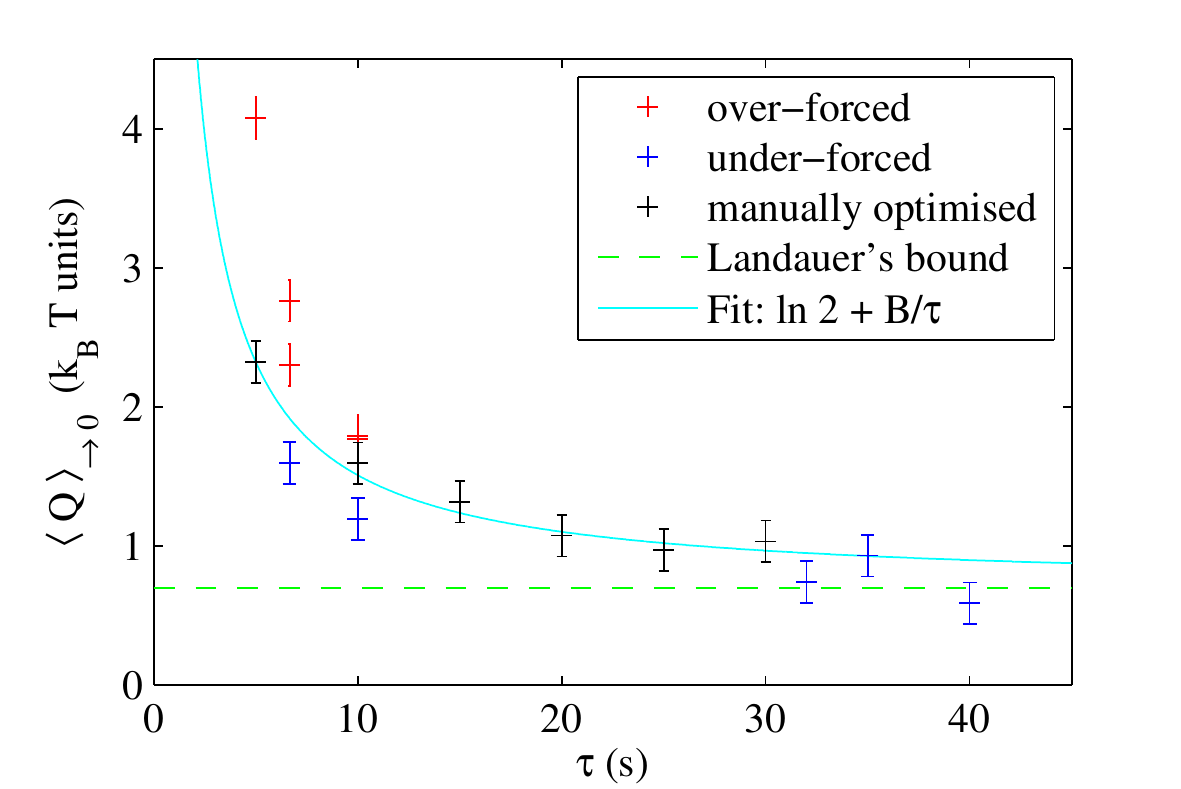}
\caption{Mean dissipated heat for several procedures, with fixed $\tau$ and different values of $f_{\text{max}}$.
 The \textcolor{red}{red} points have a force too high, and a $P_{S \, \text{force}} \geq \SI{99}{\percent}$. 
 The \textcolor{blue}{blue} points have a force too low and $\SI{91}{\percent} \leq P_{S \, \text{force}} < \SI{95}{\percent}$ 
 (except the last point which has $P_{S \, \text{force}} \approx \SI{80}{\percent}$). 
 The black points are considered to be optimised and have $\SI{95}{\percent} \leq P_{S \, \text{force}} < \SI{99}{\percent}$. 
 The errors bars are $\pm \SI{0.15}  {\kbT}$ estimated from the reproductibility of measurement with same parameters. 
 The fit $\langle Q \rangle_{\rightarrow 0} = \ln 2 + B/\tau$ is done only by considering the optimised procedures.}
\label{landauer:fig:mean_dissipated_heat}
\end{center}
\end{figure}

The Landauer's bound $k_{\text{B}}T \ln 2$ is only valid for totally efficient procedures. Thus one should theoretically look for a Landauer's bound 
corresponding to each experimental proportion of success (see equation~\ref{landauer:eq:landauer_tot}). Unfortunately the function $\ln 2 + p \ln 
(p) + (1-p) \ln (1-p)$ quickly decreases when $p$ is lower than 1. To avoid this problem, we made an approximation by computing $\langle Q 
\rangle_{\rightarrow 0}$ the mean dissipated heat for the trajectories where the memory is erased (\textit{i.e} the ones ending in state 0). We 
consider that $\langle Q \rangle_{\rightarrow 0}$ mimics the mean dissipated heat for a procedure with $\SI{100}{\percent}$ of success. This 
approximation is reasonable as long as $P_{S \, \text{force}}$ is close enough to $\SI{100}{\percent}$, because the negative contributions which 
reduce the average dissipated heat are mostly due to the rare trajectories going against the force (\textit{i.e} ending in state 1). Of course, at the 
limit where the force is equal to zero, one should find $P_{S \, \text{force}} = \SI{50}{\percent}$ and $\langle Q \rangle_{\rightarrow 0} = 0$ which 
is different from $k_{\text{B}}T \ln 2$. The mean dissipated heat for several procedures are shown in figure~\ref
{landauer:fig:mean_dissipated_heat}.

The mean dissipated heat decreases with the duration of the erasure procedure $\tau$ and approaches the Landauer's bound $k_{\text{B}}T \ln 
2$ for long times. Of course, if we compute the average on all trajectories (and not only on the ones ending in state 0) the values of the mean 
dissipated heat are smaller, but remain greater than the generalised Landauer's bound for the corresponding proportion of success $p$:
\begin{equation}
\langle Q \rangle_{\rightarrow 0} \geq \langle Q \rangle \geq k_{\text{B}}T \left[ \ln 2 + p \ln (p) + (1-p) \ln (1-p) \right]
\end{equation}
For example, the last point ($\tau = \SI{40}   {\second}$) has a proportion of success $P_{S \, \text{force}} \approx \SI{80}{\percent}$, which 
corresponds to a Landauer's bound of only $\approx \SI{0.19}    {\kbT}$, and we measure 
$\langle Q \rangle_{\rightarrow 0} = \SI{0.59}    {\kbT}$ greater than $\langle Q \rangle = \SI{0.26}    {\kbT}$.
The manually optimised procedures also seem to verify a decreasing of $\langle Q \rangle_{\rightarrow 0}$ proportional to $1/\tau$. A numerical 
least square fit $\langle Q \rangle_{\rightarrow 0} = \ln 2 + B/\tau$ is plotted in figure~\ref{landauer:fig:mean_dissipated_heat} and gives a value of 
$B = \SI{8.15}{\kbT\second}$. If we do a fit with two free parameters $\langle Q \rangle_{\rightarrow 0} = A + B/\tau$, we find $A = \SI{0.72}  {\kbT}
$ which is close to $k_{\text{B}}T \ln 2 \approx \SI{0.693}  {\kbT}$.

\begin{figure}[ht!]
        \centering
        \begin{subfigure}[c]{0.5\textwidth}
                \includegraphics[width=\textwidth]{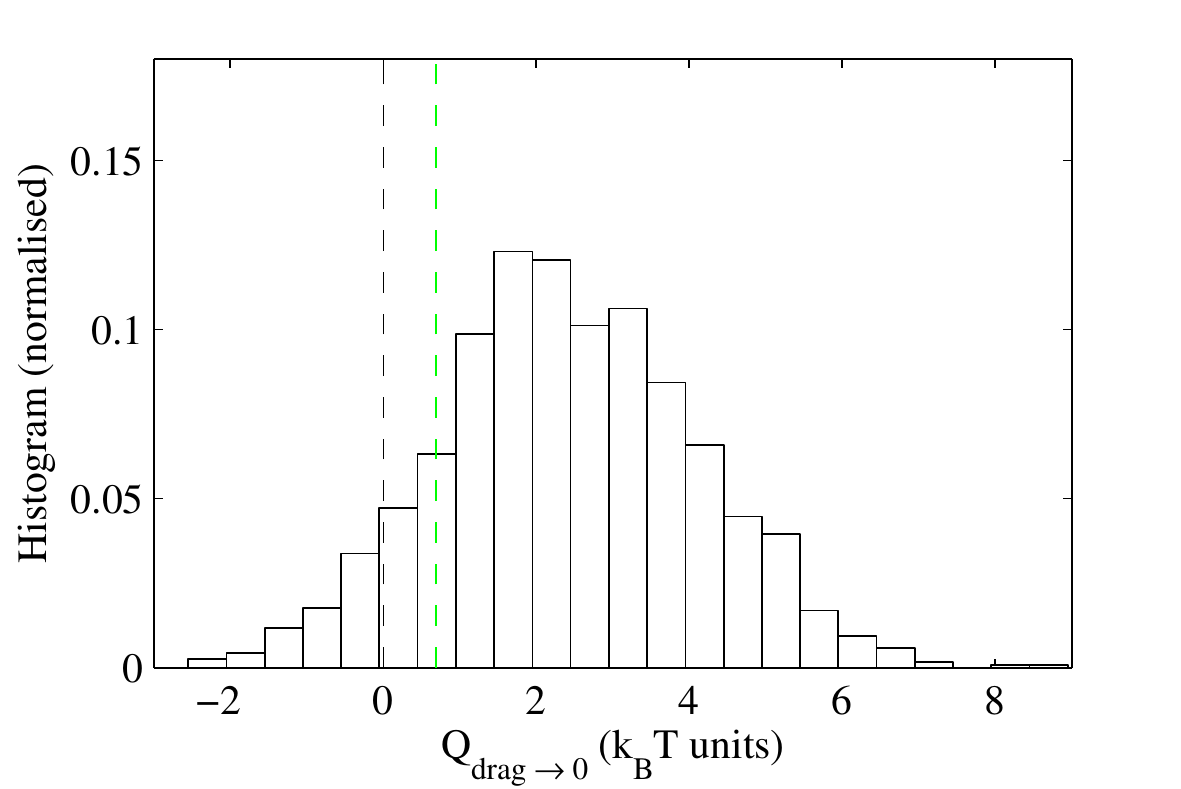}
                \caption{$1 \rightarrow 0$}
                \label{landauer:fig:hist_1vers0}
        \end{subfigure}%
        \begin{subfigure}[c]{0.5\textwidth}
                \includegraphics[width=\textwidth]{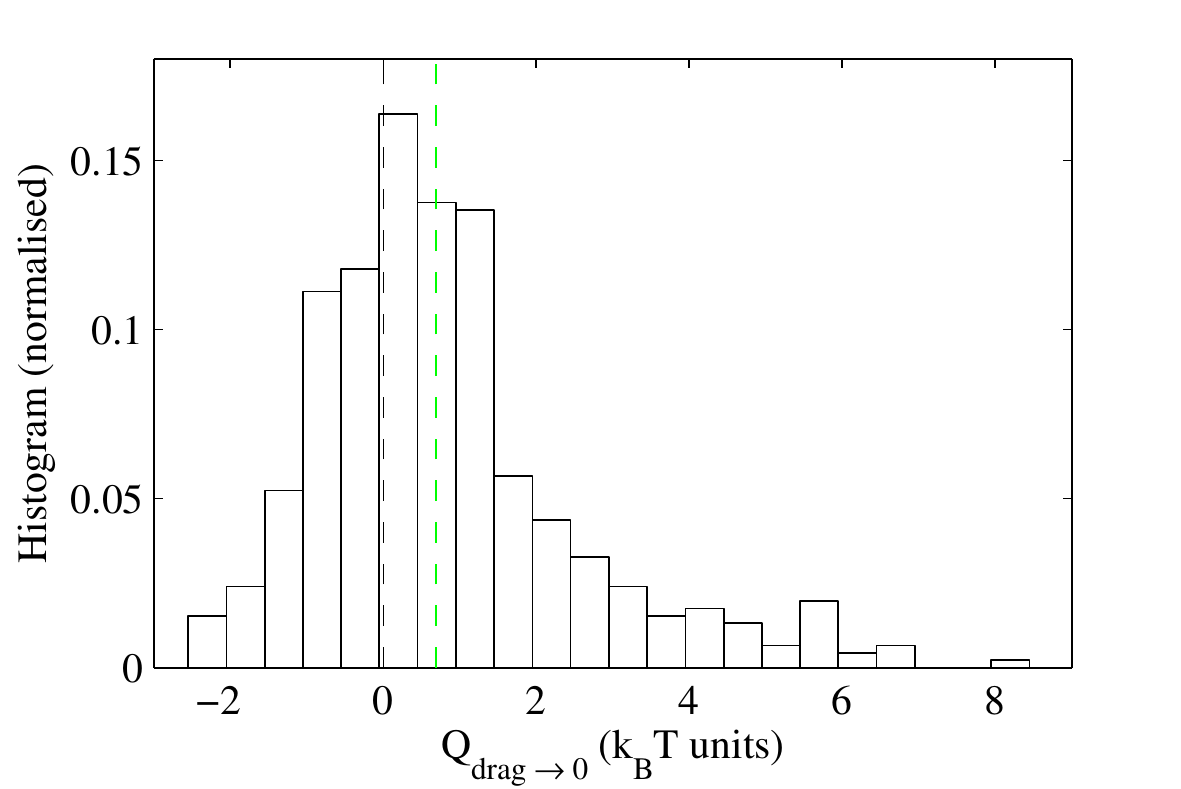}
                \caption{$0 \rightarrow 0$}
                \label{landauer:fig:hist_0vers0}
        \end{subfigure}
        \caption{Histograms of the dissipated heat $Q_{\text{drag} \, \rightarrow 0}$. (a) For one procedure going from 1 to 0 ($\tau = \SI{10}  
{\second}$ and $f_{\text{max}} = \SI{3.8e-14}{\newton}$). (b) For one procedure going from 0 to 0 
        ($\tau = \SI{5}  {\second}$ and $f_{\text{max}} = \SI{3.8e-14}{\newton}$).
         The black vertical lines indicate $Q=0$ and the green ones indicate the Landauer's bound $k_{\text{B}}T \ln 2$.}
        \label{landauer:fig:histograms}
\end{figure}

One can also look at the distribution of $Q_{\text{drag} \, \rightarrow 0}$. Histograms for procedures going from 1 to 0 and from 0 to 0 are shown 
in figure~\ref{landauer:fig:histograms}. The statistics are not sufficient to conclude on the exact shape of the distribution, but as expected, there is 
more heat dissipated when the particle has to jump from state 1 to state 0 than when it stays in state 0. It is also noticeable that a fraction of the 
trajectories always dissipate less heat than the Landauer's bound, and that some of them even have a negative dissipated heat. We are able to 
approach the Landauer's bound in average thanks to those trajectories where the thermal fluctuations help us to erase the information without 
dissipating heat.

\newpage

\section{Integrated Fluctuation Theorem applied on information erasure procedure}

We have shown that the mean dissipated heat for an information erasure procedure applied on a 1 bit memory system approaches the 
Landauer's bound for quasi-static transformations. One may wonder if we can have direct access to the variation of free-energy between the 
initial and the final state of the system, which is directly linked to the variation of the system's entropy.

\subsection{Computing the stochastic work}

To answer to this question it seems natural to use the Integrated Fluctuation Theorem called the Jarzinsky equality~\cite{Jarzynski1997} which 
allows one to compute the free energy difference between two states of a system, in contact with a heat bath at temperature $T$. When such a 
system is driven from an equilibrium state A  to a state B through any continuous procedure, the Jarzynski equality links the stochastic work $W_
{\text{st}}$  received by the system during the procedure to the free energy difference $\Delta F = F_{B}-F_{A}$ between the two states:
\begin{equation}
\left\langle \mathrm{e}^{- \beta W_{\text{st}}} \right\rangle = \mathrm{e}^{-\beta \Delta F}
\end{equation}
Where  $\left\langle . \right\rangle$ denotes the ensemble average over all possible trajectories, and $\beta = \frac{1}{k_{\text{B}}T}$.

For a colloidal particle confined in one spatial dimension and submitted to a conservative potential $V(x,\lambda)$, where $\lambda=\lambda(t)$ 
is a time-dependent external parameter, the stochastic work received by the system is defined by~\cite{SeifertReview}:
\begin{equation}
W_{\text{st}}[x(t)]=\int_{0}^{t} \frac{\partial V}{\partial \lambda} \dot{\lambda} \, \mathrm{d} t^{\prime}
\end{equation}
Here the potential is made by the double-well and the tilting drag force\footnote{Since we are in one dimension, any external force can be written 
as the gradient of a global potential.}
\begin{equation}
V(x,\lambda) =  U_0(x,I(t)) - f(t) x
\end{equation}
and we have two control parameters: $I(t)$ the intensity of the laser and $f(t)$ the amplitude of the drag force.

Once again, we can separate two contributions: one coming from the lowering and rising of the barrier, and one coming from the applied external 
drag force. We again consider that the lowering and rising of the barrier should not modify the free-energy of the system, and that the main 
contribution is due to the drag force. Thus:
\begin{equation}
W_{\text{st}} = \int_{T_{\text{low}}}^{T_{\text{low}}+\tau} - \dot{f}  x  \, \mathrm{d} t^{\prime}
\end{equation}
Noting that $f(t=T_{\text{low}})=0=f(t=T_{\text{low}}+\tau)$, it follows from an integration by parts that the stochastic work is equal to the heat 
dissipated by the drag force:
\begin{equation}
W_{\text{st}} = \int_{T_{\text{low}}}^{T_{\text{low}}+\tau} - \dot{f} x  \, \mathrm{d} t^{\prime} = \int_{T_{\text{low}}}^{T_{\text{low}}+\tau} f \dot{x} \, 
\mathrm{d} t^{\prime} = Q_{\text{drag}}
\end{equation}
The two integrals have been calculated experimentally for all the trajectories of all the procedures tested and it was verified that the difference 
between the two quantity is completely negligible. In the following parts, we write $W_{\text{st}}$ for theoretic calculations, and $Q_{\text{drag}}$ 
when we apply the calculations to our experimental data.

\subsection{Interpreting the free-energy difference}

Since the memory erasure procedure is made in a cyclic way (which implies $\Delta U = 0$) and $\Delta S = - k_{\text{B}} \ln 2$ it is natural to 
await $\Delta F = k_{\text{B}}T \ln 2$. But the $\Delta F$ that appears in the Jarsynski equality is the difference between the free energy of the 
system in the initial state (which is at equilibrium) and the equilibrium state corresponding to the final value of the control parameter:
\begin{equation}
\Delta F_{\text{Jarzynski}} = F(\lambda(t_{\text{final}}))-F(\lambda(t_{\text{initial}}))
\end{equation}
Because the height of the barrier is always finite there is no change in the equilibrium free energy of the system between the beginning and the 
end of our procedure. Then $\Delta F_{\text{Jarzynski}} = 0$, and we await $\left\langle \mathrm{e}^{- \beta W_{\text{st}}} \right\rangle = 1$, which 
is not very interesting. 

Nevertheless it has been shown~\cite{Jarzynski2009} that, when there is a difference between the actual state of the system (described by the 
phase-space density $\rho _{t}$) and the equilibrium state (described by $\rho ^{\text{eq}}_{t}$), the Jarzynski equality can be modified:
\begin{equation}
\left\langle \mathrm{e}^{- \beta W_{\text{st}}(t)} \right\rangle _{(x,t)} = \frac{\rho ^{\text{eq}}(x,\lambda (t))}{\rho (x,t)} \mathrm{e}^{-\beta \Delta F_
{\text{Jarzynski}}(t)}
\label{landauer:eq:jarzynski}
\end{equation}
Where $\left\langle . \right\rangle _{(x,t)}$ is the mean on all the trajectories that pass through $x$ at $t$.

In our procedure, selecting the trajectories where the information is actually erased is equi\-valent to fix the position $x$ to the chosen final well 
(state 0 corresponds to $x < 0$) at the time $t=T_{\text{low}}+\tau$. It follows that $\rho (x<0,T_{\text{low}}+\tau)$ is directly $P_{S \, \text{force}}
$, the proportion of success of the procedure, and $\rho ^{\text{eq}}(x<0,\lambda (T_{\text{low}}+\tau)) = 1/2$ since both wells have same 
probability at equilibrium\footnote{A more detailed demonstration is given in Appendix.}. Then:
\begin{equation}
\left\langle \mathrm{e}^{- \beta W_{\text{st}}(T_{\text{low}}+\tau)} \right\rangle _{\rightarrow 0} = \frac{1/2}{P_{S \, \text{force}}}
\label{landauer:eq:<>0}
\end{equation}
Similarly for the trajectories that end the procedure in the wrong well (state 1) we have:
\begin{equation}
\left\langle \mathrm{e}^{- \beta W_{\text{st}}(T_{\text{low}}+\tau)} \right\rangle _{\rightarrow 1} = \frac{1/2}{1-P_{S \, \text{force}}}
\label{landauer:eq:<>1}
\end{equation}
Taking into account the Jensen's inequality, i.e. 
$ \left\langle \mathrm{e}^{-x} \right\rangle \ge \mathrm{e}^{-\left\langle x \right\rangle}$, we find that equations~\ref{landauer:eq:<>0} and~\ref
{landauer:eq:<>1} imply:
\begin{equation}
\begin{array}{l}
\left\langle W_{\text{st}} \right\rangle_{\rightarrow 0} \geq k_{\text{B}} T \left[ \ln(2) + \ln(P_{S \, \text{force}}) \right]  \\
\left\langle W_{\text{st}} \right\rangle_{\rightarrow 1} \geq k_{\text{B}} T \left[ \ln(2) + \ln(1-P_{S \, \text{force}}) \right]
\end{array}
\end{equation}
Since that the mean stochastic work dissipated to realise the procedure is simply: 
\begin{equation}
\left\langle W_{\text{st}} \right\rangle = P_{S \, \text{force}} \left\langle W_{\text{st}} \right\rangle_{\rightarrow 0} + (1-P_{S \, \text{force}}) \left\langle 
W_{\text{st}} \right\rangle_{\rightarrow 1}
\end{equation}
it follows:
\begin{equation}
\left\langle W_{\text{st}} \right\rangle \geq k_{\text{B}} T \left[ \ln(2) + P_{S \, \text{force}} \ln(P_{S \, \text{force}}) + (1-P_{S \, \text{force}}) \ln(1-P_
{S \, \text{force}}) \right]
\end{equation} 
which is the generalization of the Landauer's bound for $P_{S \, \text{force}} < \SI{100}{\percent}$. Hence, the Jarzynski equality applied to the 
information erasure procedure allows one to find the complete Landauer's bound for the stochastic work received by the system.

\begin{figure}[ht!]
\begin{center}
\includegraphics[width=9cm]{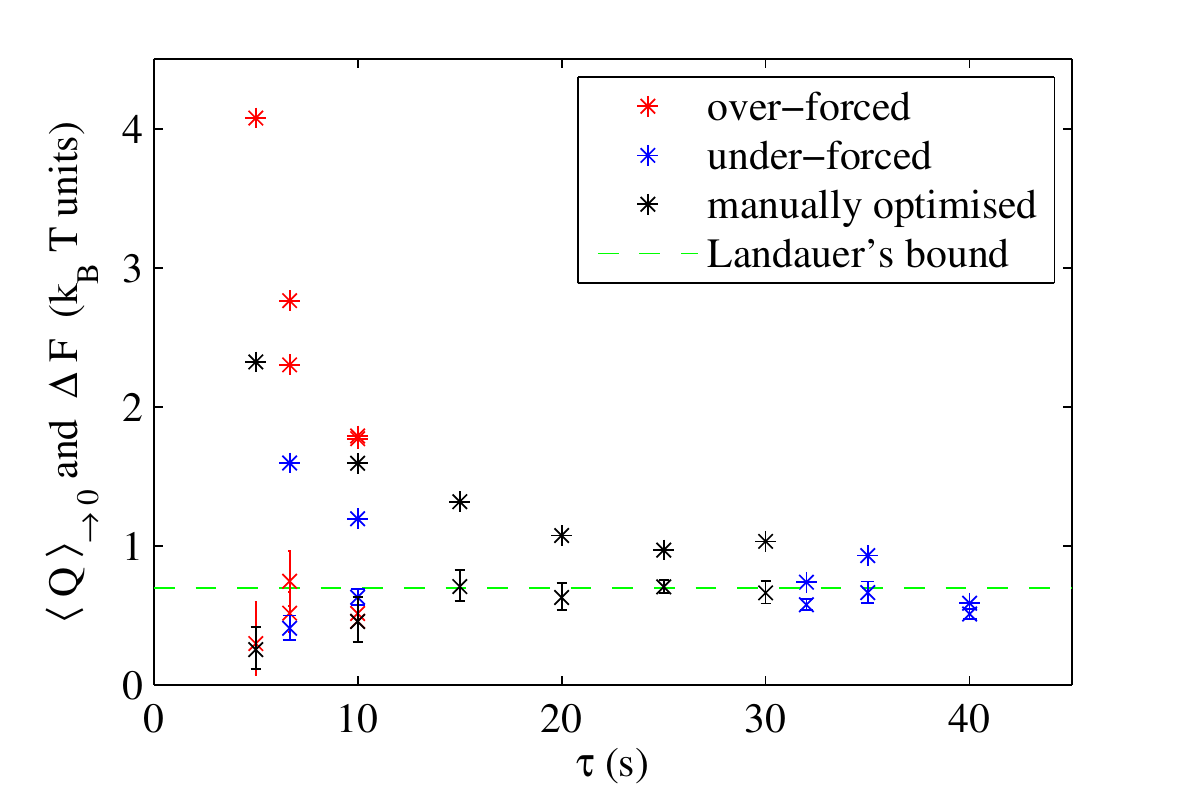}
\caption{Mean dissipated heat ($\ast$) and effective free energy difference ($\times$) for several procedures, with fixed $\tau$ and different 
values of $f_{\text{max}}$. The \textcolor{red}{red} points have a force too high, and a $P_{S \, \text{force}} \geq \SI{99}{\percent}$. The \textcolor
{blue}{blue} points have a force too low and $\SI{91}{\percent} \leq P_{S \, \text{force}} < \SI{95}{\percent}$ (except the last point which has $P_{S 
\, \text{force}} \approx \SI{80}{\percent}$). The black points are considered to be optimised and have $\SI{95}{\percent} \leq P_{S \, \text{force}} < 
\SI{99}{\percent}$.}
 \label{landauer:fig:DeltaF_jarzynski}
\end{center}
\end{figure}

Finally we experimentally compute $\Delta F_{\text{eff}}$ which is the logarithm of the exponential average of the dissipated heat for trajectories 
ending in state 0:
\begin{equation}
\Delta F_{\text{eff}} =  -\ln \left( \left\langle \mathrm{e}^{- \beta Q_{\text{drag}}} \right\rangle_{\rightarrow 0} \right) .
\end{equation}
Data are shown in figure~\ref{landauer:fig:DeltaF_jarzynski}. The error bars are estimated by computing the average on the data set with $10\%$ 
of the points randomly excluded, and taking the maximal difference in the values observed by repeating this operation $1000$ times. Except for 
the first points\footnote{We believe that the discrepancy can be explained by the fact that the values of $Q_{\text{drag} \, \rightarrow 0}$ are 
bigger and that it is more difficult to estimate correctly the exponential average in this case.} ($\tau = \SI{5}  {\second}$), the values are very close 
to $k_{\text{B}}T \ln 2$, which is in agreement with equation~\ref{landauer:eq:<>0}, since $P_{S \, \text{force}}$ is close to $\SI{100}{\percent}$. 
Hence, we retrieve the Landauer's bound for the free-energy difference, for any duration of the information erasure procedure.

Note that this result is not in contradiction with the classical Jarzynski equality, because if we average over all the trajectories (and not only the 
ones where the information is erased), we should find:
\begin{equation}
\left\langle \mathrm{e}^{- \beta W_{\text{st}}} \right\rangle = P_{S \, \text{force}} \left\langle \mathrm{e}^{- \beta W_{\text{st}}} \right\rangle_
{\rightarrow 0} + (1-P_{S \, \text{force}}) \left\langle \mathrm{e}^{- \beta W_{\text{st}}} \right\rangle_{\rightarrow 1} = 1.
\end{equation}
However, the verification of this equality is hard to do experimentally since we have very few trajectories ending in state 1, which gives us not 
enough statistics to estimate $\left\langle \mathrm{e}^{- \beta W_{\text{st}}} \right\rangle_{\rightarrow 1}$ properly.

\subsection{Separating sub-procedures}

To go further, we can also look at the two sub-procedures $1 \rightarrow 0 $ and $ 0 \rightarrow 0$ separately. To simplify calculations, we make 
here the approximation that $P_{S \, \text{force}} = \SI{100}{\percent}$.

We can compute the exponential average of each sub-procedure:
\begin{equation}
M_{1 \rightarrow 0}=\left\langle \mathrm{e}^{-\beta W_{\text{st}}} \right\rangle_{1 \rightarrow 0} \quad \textrm{and} \quad M_{0 \rightarrow 0}=\left
\langle \mathrm{e}^{-\beta W_{\text{st}}} \right\rangle_{0 \rightarrow 0}
\label{landauer:eq:MM}
\end{equation} 
For each sub-procedure taken independently the classical Jarzynski equality does not hold because the initial conditions are not correctly tested. 
Indeed selecting trajectories by their initial condition introduces a bias in the initial equilibrium distribution. But it has been shown~\cite
{Kawai2007} that for a partition of the phase-space into non-overlapping subsets $\chi_{j}$ ($j=1,...,K$) there is a detailed Jarzynski Equality :
\begin{equation}
\left\langle \mathrm{e}^{- \beta W_{\text{st}}} \right\rangle_{j} 
= \frac{\tilde{\rho}_j}{\rho_j} \left\langle \mathrm{e}^{- \beta W_{\text{st}}} \right\rangle
\label{landauer:eq:vandenbroeck}
\end{equation}
with:
\begin{equation}
\rho_j = \int_{\chi_{j}} \rho(t_{a}) \, \mathrm{d}x\mathrm{d}p ~~\mbox{and}~~ \tilde{\rho}_j = \int_{\tilde{\chi}_{j}} \tilde{\rho}(t_{a}) \, \mathrm{d}x
\mathrm{d}p
\end{equation}
where $\rho(t_{a})$ and $\tilde{\rho}(t_{a})$ are the phase-space densities of the system measured at the same intermediate but otherwise 
arbitrary point in time, in the forward and backward protocol, respectively. The backward protocol is simply the time-reverse of the forward 
protocol.

\begin{figure}[ht!]
\begin{center}
\includegraphics[width=9cm]{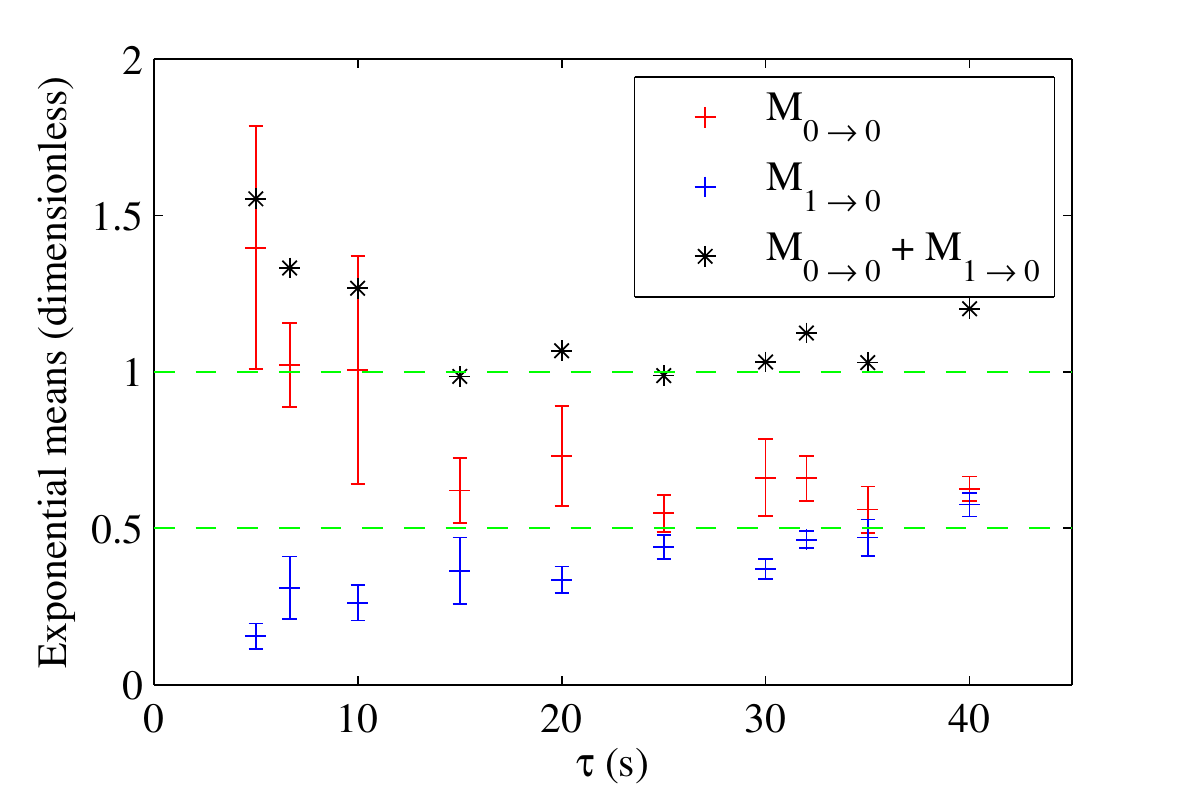}
\caption{Exponential means computed on the sub-procedures, for several parameters, with fixed $\tau$ and manually optimised values of $f_
{\text{max}}$. The errorbars are estimated by computing the exponential mean on the data set with $10\%$ of the points randomly excluded, and 
taking the maximal difference in the values observed by repeating this operation $1000$ times.}
 \label{landauer:fig:exponential_means}
\end{center}
\end{figure}

Here, we take only two subsets $j=\{0 \rightarrow 0,1 \rightarrow 0\}$, defined by the position where the bead starts, and we choose $t_{a}=T_
{\text{low}}$ the starting point of the applied force. Then we have:
\begin{equation}
M_{i \rightarrow 0} = \frac{\tilde{\rho}_{i \rightarrow 0}}{\rho_{i \rightarrow 0}} \mathrm{e}^{- \beta \Delta F_{\text{eff}}} = \frac{\tilde{P}_{0 \rightarrow 
i}}{1/2} \, \frac{1}{2}
\end{equation}
where $i =\{ 0 , 1 \} $ and $\tilde{P}_{0 \rightarrow i}$ is the probability that the system returns into its initial state $i$ under the time-reversed 
procedure (which always starts in state 0).
Finally:
\begin{equation}
M_{1 \rightarrow 0} = \tilde{P}_{0 \rightarrow 1} \quad \textrm{and} \quad M_{0 \rightarrow 0} = \tilde{P}_{0 \rightarrow 0}
\label{landauer:eq:timereversed}
\end{equation}

Experimental data are shown in figure~\ref{landauer:fig:exponential_means}. $M_{1 \rightarrow 0}$ is an increasing function of $\tau$ whereas $ 
M_{0 \rightarrow 0}$ is decreasing with $\tau$. Their sum is always close to 1 (which is the same result that gives $\Delta F_{\text{eff}} \approx k_
{\text{B}}T \ln 2$), and $M_{1 \rightarrow 0}$ is always smaller than $M_{0 \rightarrow 0}$. These observations are intuitive because the work is 
higher when the bead jumps from state 1 to 0 than when it stays in state 0, and the work is higher when $\tau$ is smaller. The interpretation of 
eq.~\ref{landauer:eq:timereversed} gives a little more information. It is indeed reasonable to think that for time-reversed procedures the probability 
of returning to state 1 is small for fast procedures and increases when $\tau$ is bigger, whereas the probability of returning to state 0 increase 
when $\tau$ is smaller\footnote{Of course $\tilde{P}_{0 \rightarrow 1}+\tilde{P}_{0 \rightarrow 0}=1.$}.

To be more quantitative one has to measure  $\tilde{P}_{0 \rightarrow 1}$ and $\tilde{P}_{0 \rightarrow 0}$, but the time-reversed procedure 
cannot be realised experimentally, because it starts with a very fast rising of the force, which cannot be reached in our experiment. Thus, we 
performed numerical simulations, where it is possible to realise the corresponding time-reversed procedure and to compute $\tilde{P}_{0 
\rightarrow 1}$ and $\tilde{P}_{0 \rightarrow 0}$. We simply integrate eq.~\ref{landauer:eq:langevin} with Euler's method, for different set of 
parameters as close as possible to the experimental ones. The Gaussian white noise is generated by the ``randn'' function from Matlab$^
{\circledR}$ (normally distributed pseudorandom numbers). For each set of parameters we repeat the numerical procedure a few thousand of 
times. Some results are shown in table~\ref{landauer:tab:simulations} (values are estimated with errorbar $\pm 0.02$):

\begin{table}[ht!]
\begin{center}
\begin{tabular}{|c|c|c|c|c|c|c|c|}
  \hline
   & & & & & & & \\
  $\tau$ (\si{s}) & $f_{\text{max}}$ (\si{fN}) & $M_{1 \rightarrow 0}$ & $\tilde{P}_{0 \rightarrow 1}$ & $M_{0 \rightarrow 0}$ & $\tilde{P}_{0 
\rightarrow 0}$ & $P_S$ (\si{\percent}) & $P_{S \, \text{force}}$ (\si{\percent}) \\
   & & & & & & & \\
  \hline
  5 & 37.7 & 0.17 & 0.16 & 0.86 & 0.84 & 97.3 & 99.8 \\
  10 & 28.3 & 0.29 & 0.28 & 0.74 & 0.72 & 96.6 & 99.3 \\
  20 & 18.9 & 0.42 & 0.41 & 0.63 & 0.59 & 94 & 97.1 \\
  30 & 18.9 & 0.45 & 0.43 & 0.59 & 0.57 & 94.4 & 97.7 \\
  \hline
\end{tabular}
\end{center}
\caption{Results for simple numerical simulations of the experimental procedure.}
\label{landauer:tab:simulations}
\end{table}

All the qualitative behaviours observed in the experimental data are retrieved, and the agreement between $M_{i \rightarrow 0}$ and $\tilde{P}_{0 
\rightarrow i}$ is correct. It was also verified that for proportions of success $< \SI{100}{\percent}$, if one takes all the trajectories, and not only the 
ones where the bead ends in the state 0, the classical Jarzynski equality is verified: $\left\langle \mathrm{e}^{- \beta W_{\text{st}}} \right\rangle = 
1$. This result means that the small fraction of trajectories where the bead ends the erasure procedure where it shouldn't, which represent 
sometimes less than $\SI{1}{\percent}$ of all trajectories, is enough to retrieve the fact that $\Delta F_{\text{Jarzynski}} = 0$.

\newpage

\section{Conclusion}

In conclusion, we have realised an experimental information erasure procedure with a 1-bit memory system, made of a micro-particle trapped in a 
double well potential with optical tweezers. The procedure uses an external drag force to reset the memory of the system in one state, \textit{i.e.} 
erase the knowledge of previous state and lose information. We measured the proportion of success of erasure procedures with different duration 
$\tau$ and amplitude of the force $f_{\text{max}}$. These data were used to manually optimised the procedure, \textit{i.e.} for different $\tau$ we 
found the lowest force which gives a good erasure of information. By varying the duration of the information erasure procedure $\tau$, we were 
able to approach the Landauer's bound $k_{\text{B}}T \ln 2$ for the mean dissipated heat by the system $\langle Q \rangle$. We have also shown 
that $\langle Q \rangle$ seems to decrease as $1/\tau$, which is in agreement with the theoretical prediction for an optimal information erasure 
procedure~\cite{Aurell2012}, and was later confirmed experimentally with a more controlled experimental system~\cite{Bechhoefer2014}.

We have computed the stochastic work received by the system during the procedure, which is in our particular case equal to the heat dissipated 
by the action of the external force. We used a modified version of the Jarzynski equality~\cite{Jarzynski2009} for systems ending in a non-
equilibrium state to retrieve the generalised Landauer's bound for any proportion of success on the mean stochastic work received by the system. 
This relation has been tested experimentally, and we have shown that the exponential average of the stochastic work, computed only on the 
trajectories where the information is actually erased, reaches the Landauer's bound for any duration of the procedure.

We also used a detailed version of the Jarzynski equality~\cite{Kawai2007} to consider each sub-procedure where the information is erased ($1 
\rightarrow 0$ and $0 \rightarrow 0$) independently. This relation allowed us to link the exponential average of stochastic work, computed only on 
a subset of the trajectories (corresponding to one of the sub-procedures), to the probability that the system returns to its initial state under a time-
reversed procedure. We have shown that the experimental data are qualitatively in agreement with this interpretation. Finally, we used some very 
simple numerical simulations of our experimental procedure to compare quantitatively the partial exponential averages to the probabilities that the 
system returns in its initial state under time-reversed procedures.

\newpage

\section*{Appendix}
\addcontentsline{toc}{section}{Appendix}
\label{landauer:sec:appendix}

Equation~\ref{landauer:eq:<>0} is obtained directly if the system is considered as a two state system, but it also holds if we consider a bead that 
can take any position in a continuous 1D double potential along the $x$-axis. We place the reference $x=0$ at the center of the double potential.

\noindent Equation~\ref{landauer:eq:jarzynski} states:
\begin{equation}
\left\langle \mathrm{e}^{- \beta W_{\text{st}}(t)} \right\rangle _{(x,t)} = \frac{\rho ^{\text{eq}}(x,\lambda (t))}{\rho (x,t)} \mathrm{e}^{-\beta \Delta F_
{\text{Jarzynski}}(t)}
\label{landauer:eq:paper}
\end{equation}
where $\left\langle . \right\rangle _{(x,t)}$ is the mean on all the trajectories that pass through $x$ at $t$.

\noindent We choose $t= T_{\text{low}} + \tau$ the ending time of the procedure, and we will not anymore write the explicit dependence upon $t$ 
since it's always the same chosen time.
We recall that $\Delta F_{\text{Jarzynski}} = 0$ at $t = T_{\text{low}} + \tau$ for our procedure.

\noindent We define the proportion of success, which is the probability that the bead ends its trajectory in the left half-space $x<0$:
\begin{equation}
P_{S \, \text{force}} = \rho(x < 0) = \int_{-\infty}^{0} \mathrm{d}x \,  \rho (x)
\end{equation}

\noindent The conditional mean is given by:
\begin{equation}
\left\langle \mathrm{e}^{- \beta W_{\text{st}}} \right\rangle _{x} = \int \mathrm{d}W_{\text{st}} \,  \rho (W_{\text{st}} | x) \mathrm{e}^{- \beta W_{\text
{st}}}
\label{landauer:eq:defmean}
\end{equation}
where $\rho (W_{\text{st}} | x)$ is the conditional density of probability of having the value $W_{\text{st}}$ for the work, knowing that the trajectory 
goes through $x$ at the chosen time $\tau$.

\noindent We recall from probability properties that:
\begin{equation}
\rho (W_{\text{st}} | x) = \frac{\rho (W_{\text{st}},x)}{\rho (x)}
\label{landauer:eq:condprob}
\end{equation}
where $\rho (W_{\text{st}},x)$ is the joint density of probability of the value $W_{\text{st}}$ of the work and the position $x$ through which the 
trajectory goes at the chosen time $\tau$.

\noindent We also recall: 
\begin{equation}
\rho (W_{\text{st}} | x < 0) = \frac{\int_{-\infty}^{0} \mathrm{d}x \, \rho (W_{\text{st}},x)}{\int_{-\infty}^{0} \mathrm{d}x \, \rho (x)}=\frac{\int_{-\infty}^{0} 
\mathrm{d}x \, \rho (W_{\text{st}},x)}{P_{S \, \text{force}}}
\label{landauer:eq:condprobint}
\end{equation}

\noindent Then by multiplying equation~\ref{landauer:eq:paper} by $\rho (x)$ and integrating over the left half-space $x < 0$ we have:
\begin{equation}
\int_{-\infty}^{0} \mathrm{d}x \, \rho (x) \left\langle \mathrm{e}^{- \beta W_{\text{st}}} \right\rangle _{x} = \int_{-\infty}^{0} \mathrm{d}x \, \rho ^{\text
{eq}} (x)
\end{equation}

\noindent Since the double potential is symmetric $\int_{-\infty}^{0} \mathrm{d}x \, \rho ^{\text{eq}} (x) = \frac{1}{2}$.

\noindent By applying definition~\ref{landauer:eq:defmean} and equality~\ref{landauer:eq:condprob}, it follows:
\begin{equation}
\int_{-\infty}^{0} \mathrm{d}x \, \int \mathrm{d}W_{\text{st}} \, \rho (W_{\text{st}} , x) \mathrm{e}^{- \beta W_{\text{st}}} = \frac{1}{2}
\end{equation}

\noindent Then using equality~\ref{landauer:eq:condprobint}:
\begin{equation}
P_{S \, \text{force}} \int \mathrm{d}W_{\text{st}} \, \rho (W_{\text{st}} | x < 0) \mathrm{e}^{- \beta W_{\text{st}}} = \frac{1}{2}
\end{equation}

\noindent Finally we obtain:
\begin{equation}
\left\langle \mathrm{e}^{- \beta W_{\text{st}}} \right\rangle _{x(T_{\text{low}}+\tau)<0} = \frac{1/2}{P_{S \, \text{force}}}
\end{equation}
which is equation~\ref{landauer:eq:<>0} of the main text.

\newpage

\bibliographystyle{plain}
\bibliography{biblio}
\addcontentsline{toc}{section}{Bibliography}

\end{document}